\documentclass[11pt]{article}
\usepackage{longtable,supertabular}
\usepackage{amsmath}
\usepackage{amssymb}
\usepackage{latexsym}
\usepackage{cite}

\title{\bf Star Product PIR Schemes with Colluding Servers over Small Fields}
\author{Hao Chen
  \thanks{Hao Chen is with the College of Information Science and Technology/Cyber Security, Jinan University, Guangzhou, Guangdong Province, 510632, China, haochen@jnu.edu.cn. The research of Hao Chen was supported by NSFC Grant 62032009.}}

\begin{document}

\maketitle
\begin{abstract}
Private Information Retrieval (PIR) was first proposed by B. Chor, O. Goldreich, E. Kushilevitz and M. Sudan in their 1995 FOCS paper. For MDS coded distributed storage system private information retrieval was proposed and the capacity of PIR schemes for MDS coded distributed storage was studied. Star product PIR schemes from general coded distributed storage system with colluding servers were constructed over general finite fields. These star product schemes has no restriction on the sizes of fields and can be constructed for coded distributed storage across large number of servers. In this paper we first propose and prove the Singleton type upper bound on the storage rate, ratio of colluding servers and the retrieval rate of the star product PIR schemes. Secondly star product PIR schemes for coded distributed storage from algebraic geometry (AG) codes are analysed. We prove that when the number of the servers goes to the infinity, star product PIR schemes with colluding servers for AG-coded distributed storage have parameters closing to the Singleton type upper bound if the field is large. Comparing with the star product PIR schemes for Reed-Solomon coded and Reed-Muller coded distributed storage we show that PIR schemes with colluding servers for AG coded distributed storage have their performance advantages. AG-code based star product PIR schemes with colluding, Byzantine and unresponsive servers are discussed. $q$-ary cyclic code based star product PIR schemes for replicated data storage are also studied. When the storage code is the Reed-Muller code, the best choice of the retrieval code is not always the Reed-Muller code.\\

\end{abstract}

\section{Introduction and Preliminaries}

The Hamming weight $wt({\bf a})$ of a vector ${\bf a} \in {\bf F}_q^n$ is the number of non-zero coordinate positions. The Hamming distance $d({\bf a}, {\bf b})=wt({\bf a}-{\bf b})$ between two vectors ${\bf a}$ and ${\bf b}$ is the Hamming weight of ${\bf a}-{\bf b}$, that is, the number of coordinate positions where ${\bf a}$ and ${\bf b}$ are different. The Hamming distance of a code ${\bf C} \subset {\bf F}_q^n$, $$d({\bf C})=\min_{{\bf a} \neq {\bf b}} \{d({\bf a}, {\bf b}),  {\bf a} \in {\bf C}, {\bf b} \in {\bf C} \},$$  is the minimum of Hamming distances $d({\bf a}, {\bf b})$ between any two different codewords in ${\bf C}$. The minimum Hamming distance of  a linear code is its minimum Hamming weight.  For a linear $[n, k, d]_q$ code, the Singleton bound asserts $d \leq n-k+1$. When the equality holds, this code is an MDS code. Reed-Solomon codes introduced in \cite{RS} are well-known MDS codes. We refer to \cite{HP} for the theory of Hamming error-correcting codes. Let $\sigma \in {\bf S}_n$ be a permutation of $n$ coordinate positions, where ${\bf S}_n$ is the order $n!$ group of all permutations of $n$ symbols, if $\sigma({\bf C})={\bf C}$, $\sigma$ is an automorphism of this code. The subgroup of all automorphisms of this code is denoted by $Aut({\bf C})$. The code is called transitive if $Aut({\bf C})$ acts transitively at the set of $n$ symbols, that is, if for any two coordinate position $u, v \in [n]$, there is an automorphism $\sigma$ of ${\bf C}$ such that $\sigma(u)=v$.\\

The Euclidean inner product on ${\bf F}_q^n$ is defined by $$<{\bf x}, {\bf y}>=\Sigma_{i=1}^n x_iy_i,$$ where ${\bf x}=(x_1, \ldots, x_n)$ and ${\bf y}=(y_1, \ldots, y_n)$. The dual of a linear code ${\bf C}\subset {\bf F}_q^n$ is $${\bf C}^{\perp}=\{{\bf c} \in {\bf F}_q^n: <{\bf c}, {\bf y}>=0, \forall {\bf y} \in {\bf C}\}.$$  The minimum distance of the Euclidean dual is called the dual distance and is denoted by $d^{\perp}$. The componentwise product of $t$ vectors ${\bf x}_j=(x_{j,1}, \ldots, x_{j, n}) \in {\bf F}_q^n$, $j=1, \ldots, t$, is ${\bf x}_1 \star \cdots \star {\bf x}_t=(x_{1,1} \cdots x_{t,1}, \ldots, x_{1, n} \cdots x_{t,n}) \in {\bf F}_q^n$. The componentwise product of linear codes ${\bf C}_1, \ldots, {\bf C}_t$ in ${\bf F}_q^n$ is defined by $${\bf C}_1\star \cdots \star {\bf C}_t=\Sigma_{ {\bf c}_i \in {\bf C}_i} {\bf F}_q {\bf c}_1 \star \cdots \star {\bf c}_t, $$ we refer to \cite{Randri} for the study of this product. If ${\bf C}_1 \subset {\bf F}_q^n$ is a linear $[n, 1, n]_q$ code, we say two linear codes ${\bf C}$ and ${\bf C} \star {\bf C}_1$ are equivalent codes.\\

We recall the coded distributed storage system from a general linear code ${\bf C} \subset {\bf F}_{q^b}^n$ as in \cite{Freij,Freij1,Taje}. The coded distributed data storage or coded database can be used to resist failed servers when data are reading from servers. If $m$ files ${\bf x}_1, \ldots, {\bf x}_m \in {\bf F}_q^{b \times k}$ need to be stored and download at $n$ servers. Each ${\bf x}_i=(x_1^i, \ldots, x_k^i)$ can be considered as a length $k$ vector in ${\bf F}_{q^b}$ for $i=1, \ldots, m$. The files can be considered as a $m \times k$ matrix ${\bf X}=(x_j^i)_{1 \leq i \leq m, 1 \leq j \leq k}$ with entries in ${\bf F}_{q^b}$. Let ${\bf G}({\bf C})$ be a generator matrix of the linear code ${\bf C}$ with $n$ columns ${\bf g}_1, \ldots, {\bf g}_n \in {\bf F}_{q^b}^k$. Then a length $m$ vector $(<{\bf x}_1, {\bf g}_i>, \ldots, <{\bf x}_m, {\bf g}_i>) \in {\bf F}_{q^b}^m$ is stored in the $i$-th server. When the files are downloaded from servers, then if at most $d({\bf C})-1$ servers are failed, then files can be recovered. Actually in any $n-d({\bf C})+1$ columns in ${\bf g}_1, \ldots, {\bf g}_n$, there are $k$ linear independent columns, otherwise there is a weight $d({\bf C})-1$ codeword in ${\bf C}$. We call $f=\frac{d({\bf C})-1}{n}$ the ratio of tolerated failed servers of this coded distributed storage system from a general linear code ${\bf C}$. The storage rate of this distributed storage system is $R_{storage}=\frac{bk}{nb}=\frac{k}{n}$. This code is called the storage code. When $\dim({\bf C})=1$ this is the replicated data storage. From the Singleton bound $$R_{storage}+f\leq 1.$$  Hence if the ratio of tolerated failed servers is high the storage rate is low. When the number $n_i$ of servers go to the infinity and a sequence of linear codes ${\bf C}_1, {\bf C}_2, \ldots, {\bf C}_i, \ldots$ of lengths $n_1, n_2, \ldots, n_i, \ldots, $ are used in coded distributed storage systems,  we call $$R_{asym, storage}=\lim\frac{\dim({\bf C}_i)}{n_i}$$ the asymptotic storage rate.\\

Private information retrieval was proposed in \cite{Chor,Chor1}. In the classical model data ${\bf x}=(x^1, \ldots, x^m) \in {\bf F}_2^m$ is stored and the user want to retrieval a single bit $x^i$ without revealing any information about the index $i$. The retrieval rate of an PIR scheme is the ratio of the gained information over downloaded information, while uploaded costs of the requests are ignored. PIR scheme for coded distributed storage of data was proposed in \cite{Shah,Blackburn}. All files ${\bf x}_1, \ldots, {\bf x}_m \in {\bf F}_q^{b \times k}$ are stored over servers according to a storage code ${\bf C}$ as above. The coded distributed data storage can be used to help to recover the data in case of server failure. PIR scheme with $t$ colluding servers is a protocol in which the user want to retrieval the  file ${\bf x}_w$ without revealing any information of the index $w$, while any $\leq t$ servers can share their quaries from the user. In previous papers \cite{Banawan,Freij,SunJafar,SunJafar1,Zhou,Zhang} and references therein only PIR schemes for MDS coded distributed data (or replicated data) storage are considered. The MDS coded distributed storage system is maximally robust against server failure. For these coded distributed storage, the files are in ${\bf F}_q^{b \times k}$ with the condition $q^b \geq n$. That is, each file has at least $log_2 n$ bits. Obviously this is not reasonable for practice. Moreover since MDS codes over ${\bf F}_{q^b}$ satisfying $q^b\geq n$ are used, the download complexity and computational complexity of each server are lower bounded by a $log_2n$ factor, since a large field has to be used. These schemes are obviously not suitable when files with bounded length of bits are stored across large number of servers. Practical PIR schemes with colluding servers for coded databases from general linear codes over small fields were constructed and studied in \cite{Freij,Freij1,LKH,Taje}. We refer to \cite{HHW,BL20,YKPB} for computational PIR schemes, \cite{Fazeli} for PIR codes and \cite{Wang,Taje1,Saarela} for PIR schemes with colluding, Byzantine and unresponsive servers.\\

In \cite{Freij,Lin} star product PIR schemes with $t$-colluding servers for coded distributed storage from ${\bf C} \subset {\bf F}_q^n$ was constructed from another retrieval linear code ${\bf D} \subset {\bf F}_q^n$. The star product PIR schemes were studied further in \cite{Freij1,Taje,Saarela}. The start product PIR scheme has the retrieval rate $$R_{retrieval}=\frac{d({\bf C} \star {\bf D})-1}{n}$$ for general linear codes ${\bf C}$ and ${\bf D}$, see Theorem 2 in page 2111of  \cite{Freij1}. When both ${\bf C}$ and ${\bf C} \star {\bf D}$ are transitive the rate can be improved to $$R_{retrieval}=\frac{\dim(({\bf C} \star {\bf D})^{\perp})}{n},$$ see Theorem 4 in page 2113 of \cite{Freij1}.  It is clear that $$d({\bf C} \star {\bf D}) -1 \leq \dim(({\bf C} \star {\bf D})^{\perp})$$ from the Singleton bound for the linear code ${\bf C} \star {\bf D}$. Their PIR schemes protect against $d^{\perp}({\bf D})-1$ colluding servers. This is called a $(t=d^{\perp}({\bf D})-1)$-privacy PIR scheme. When the number $n_i$ of servers goes to the infinity, we call the limit $R_{asym, retrieval}=\lim R_i$ of the retrieval rates the asymptotic retrieval rate and denoted by $R_{asym, retrieval}$. The limit of the ratio of colluding servers $t_{asym}=\lim \frac{t_i}{n_i}$ is called the asymptotic colluding ratio. In star product PIR schemes the number $m$ of stored files is independent of the construction.\\

{\bf Main result 1 (Singleton type upper bound).} {\em Let ${\bf C}$ and ${\bf D}$ be two linear codes in ${\bf F}_q^n$.\\
1) If $\dim({\bf C})+d^{\perp}({\bf D}) \geq n+2$, then $d({\bf C} \star {\bf D})=1$;\\
2) Otherwise $$\dim(({\bf C} \star {\bf D})^{\perp})+d^{\perp}({\bf D}) \leq n+2-\dim({\bf C}).$$
3) If  $d^{\perp}({\bf C})+d^{\perp}({\bf D}) \geq n+3$, then  $d({\bf C} \star {\bf D})=1$.\\

 Hence the storage rate, retrieval rate and the ratio of colluding servers have to satisfy $R_{retrieval}+t \leq 1-R_{storage}+\frac{2}{n}$ for a nonzero retrieval rate star product PIR scheme with colluding servers. Asymptotically we have $$R_{asym, retrieval}+t_{asym} \leq 1-R_{asym, storage}.$$}\\

The PIR scheme with colluding servers from Reed-Muller codes and weighted lifted Reed-Muller codes were studied in \cite{Freij1,Saarela,LN21,Lava19}. These $t$-privacy PIR for coded database can be defined over the smallest field ${\bf F}_2$ and have good performance. The PIR scheme with colluding servers from the generalized Reed-Solomon codes were studied in \cite{Freij}. These $t$-privacy PIR for MDS-coded database can only be defined over the field ${\bf F}_q^{b \times k}$ satisfying $n \leq q^b$, that is, each file has at least $log_2n$ bits. The comparison with PIR scheme against colluding servers for AG-coded databases will be given in Section 4.\\

We will construct $t_i$-privacy PIR schemes from AG coded distributed storage as in the following result, in which the sequence of AG codes attaining the Tsfasman-Vl\'{a}dut-Zink bound is used.\\

{\bf Main result 2.} {\em Let ${\bf F}_{q^2}$ be a fixed base field. Let $R_1$ and $R_2$ be two positive real numbers satisfying $R_1+R_2 \geq \frac{1}{2}$. Then we have a sequence of AG-coded distributed storage with the asymptotic storage rate $$R_{asym, storage}=R_1,$$  $t_i$-privacy PIR schemes with the asymptotic retrieval rate $$R_{asym, retrieval}=1-R_1-R_2-\frac{2}{q-1}$$ and asymptotic colluding ratio $$t_{asym} \geq R_2-\frac{1}{q-1}.$$}\\

From the above construction it is obvious that $$R_{asym, retrieval}+t_{asym}=1-R_1-R_2-\frac{2}{q-1}+R_2-\frac{1}{q-1}=1-R_{asym, storage}-\frac{3}{q-1}.$$  Actually using the transitive AG codes PIR scheme against colluding servers has the improvement $\frac{1}{q-1}$. When the base field is large it is clear that PIR schemes for AG-coded distributed storage have their parameters closing to the Singleton type upper bound. The construction of PIR schemes for AG-coded distributed storage with colluding servers over a fixed base field proposed in this paper is parallel to the construction of algebraic geometric secret sharing proposed in \cite{Chen}, which is a natural generalization of the classical Shamir threshold secret sharing from MDS codes. As initial examples $q$-ary cyclic code based star product PIR schemes with colluding servers for replicated data storage are studied. Parameters of these PIR schemes are calculated based on the paper \cite{GDL,LDL,LiuLi}.\\

\section{Singleton type upper bound}

In this section we prove the Singleton type upper bound from the results in the beautiful paper \cite{Randri}.\\

{\bf Theorem 2.1.} {\em \em Let ${\bf C}$ and ${\bf D}$ be two linear codes in ${\bf F}_q^n$.\\
1) If $\dim({\bf C})+d^{\perp}({\bf D}) \geq n+2$, then $d({\bf C} \star {\bf D})=1$;\\
2) Otherwise $$\dim(({\bf C} \star {\bf D})^{\perp})+d^{\perp}({\bf D})) \leq n+2-\dim({\bf C}).$$
3) If  $d^{\perp}({\bf C})+d^{\perp}({\bf D}) \geq n+3$, then  $d({\bf C} \star {\bf D})=1$.\\

Then $R_{retrieval}+t \leq 1-R_{storage}+\frac{2}{n}$ holds for a nonzero retrieval rate start product PIR schemes with colluding servers. Asymptotically we have $$R_{asym, retrieval}+t_{asym} \leq 1-R_{asym, storage}.$$}\\

{\bf Proof.} From $$\dim({\bf C})+d^{\perp}({\bf D}) \geq n+2,$$ we have $$\dim({\bf C})+\dim({\bf D}) \geq \dim({\bf C})+d^{\perp}({\bf D})-1\geq n+1.$$ Then $n-\dim({\bf C})-\dim({\bf D})+2 \leq 1$. From Proposition 5 in \cite{Randri} we have $d({\bf C} \star {\bf D})=1$. The rate of PIR in Theorem 2 of \cite{Freij1} is zero. For PIR in Theorem 4 in \cite{Freij1}, transitivity of the code ${\bf C} \star {\bf D}$ implies that ${\bf C} \star {\bf D}$ has a weight one codeword supported at each coordinate position. Hence ${\bf C} \star {\bf D}={\bf F}_q^n$ and the rate of PIR in Theorem 4 is also zero.\\

Otherwise $$\dim({\bf C})+d^{\perp}({\bf D})-2 \leq n,$$ from Lemma 6 in \cite{Randri} we have $$\dim({\bf C} \star {\bf D}) \geq \dim({\bf C})+d^{\perp}({\bf D})-2.$$ Therefore $$n-\dim({\bf C} \star {\bf D}) +d^{\perp}({\bf  D}) \leq n-\dim({\bf C})+2.$$ The conclusion 2 is proved.\\

If $$d^{\perp}({\bf C})+d^{\perp}({\bf D}) \geq n+3,$$ then $$\dim({\bf C})+\dim({\bf D}) \geq d^{\perp}({\bf C})+d^{\perp}({\bf D})-2 \geq n+1,$$ from Singleton bound for two linear codes ${\bf C}^{\perp}$ and ${\bf D}^{\perp}$. From Proposition 5 in \cite{Randri} again, we have $d({\bf C} \star {\bf D})=1$. Therefore the rate of PIR schemes in Theorem 2 and 4  in \cite{Freij1} is zero. The conclusion is proved.\\

\section{PIR schemes for AG-coded databases over a fixed field.}

Algebraic geometry codes are suitable to be used to construct coded distributed storage, since they are natural generalizations of Reed-Solomon codes. More importantly, algebraic geometry codes with arbitrary long length can be defined over a fixed field. Therefore secret sharing over ${\bf F}_q$ with $n>q$ players corresponding to rational points of the curve can be introduced in \cite{Chen}. If the files have a fixed long length of bits and the number of the servers in the distributed storage is large, AG-coded distributed storage is more suitable.\\

Let ${\bf F}_q$ be an arbitrary finite field, $P_1,\ldots,P_n$ be $n \leq q$ elements in ${\bf F}_q$. The Reed-Solomon code  $RS(n,k)$ is defined by $$RS(n,k)=\{(f(P_1),\ldots,f(P_n)): f \in {\bf F}_q[x],\deg(f) \leq k-1\}.$$ This is an MDS $[n,k,n-k+1]_q$ linear code attaining the Singleton bound $d \leq n-k+1$ since  a degree $\deg(f) \leq k-1$ nonzero polynomial has at most $k-1$ roots. PIR schemes studied in \cite{Banawan,SunJafar,SunJafar1,Freij,Zhang,Zhou} are for MDS coded databases. They are based on the generalized Reed-Solomon codes.\\

Let ${\bf X}$ be an absolutely irreducible, smooth and genus $g$ curve defined over ${\bf F}_q$. Let $P_1,\ldots,P_n$ be $n$ distinct rational points of ${\bf X}$ over ${\bf F}_q$. Let ${\bf G}$ be a rational divisor over ${\bf F}_q$ of degree $\deg({\bf G})$ satisfying $2g-2 <\deg({\bf G})<n$ and $$support({\bf G}) \bigcap {\bf P}=\emptyset.$$ Let ${\bf L}({\bf G})$ be the function space associated with the divisor ${\bf G}$. The algebraic geometry code (functional code) associated with ${\bf G}$, $P_1,\ldots,P_n$ is defined by $${\bf C}(P_1,\ldots,P_n, {\bf G}, {\bf X})=\{(f(P_1),\ldots,f(P_n)): f \in {\bf L}({\bf G})\}.$$ The dimension of this code is $$k=\deg({\bf G})-g+1$$ follows from the Riemann-Roch Theorem. The minimum Hamming distance is $$d \geq n-\deg({\bf G}).$$ An divisor ${\bf G}=\Sigma m_i G_i$ where $m_i \geq 0$ is called an effective divisor. Algebraic-geometric residual code ${\bf C}_{\Omega}(P_1, \ldots, P_n, {\bf G}, {\bf X})$ with the dimension $k=n-m+g-1$ and minimum Hamming distance $d \geq m-2g+2$ can be defined, we refer to \cite{HP,TV} for the detail. It is the dual code of the functional code of the dimension $m-g+1$.\\

The Reed-Solomon codes are just the algebraic-geometric codes over the genus $0$ curve. One achievement of the theory of algebraic-geometric codes is the sequence of algebraic-geometric codes over ${\bf F}_{q^2}$ satisfying the Tsfasman-Vl\'{a}dut-Zink bound $$R+\delta \geq 1-\frac{1}{q-1},$$ which is exceeding the Gilbert-Varshamov bound when $q \geq 7$.  We refer to \cite{HP} Chapter 13 and \cite{TV} for the detail. In \cite{SAKS} a low-complexity polynomial time algorithm was given to construct AG codes attaining the Tsfasman-Vl\'{a}dut-Zink bound. Therefore these AG codes can be used in star product constructions.\\

\subsection{Colluding servers}

Let ${\bf X}$ be an absolutely irreducible, smooth, and  genus $g$ curve defined over ${\bf F}_q$. Let ${\bf C}={\bf C}(P_1, \ldots, P_n, {\bf G}_1, {\bf X})$ and ${\bf D}={\bf C}(P_1, \ldots, P_n, {\bf G}_2, {\bf X})$ be two AG function codes associated with two divisors $$2g-2 <\deg({\bf G}_i)<n,$$ for $i=1,2$, and $$\deg({\bf G}_1)+\deg({\bf G}_2)<n.$$ Then the storage rate of the ${\bf C}$-coded distributed storage system is $\frac{\deg ({\bf G}_1)-g+1}{n}$ and the ratio of tolerated failed servers is $f=\frac{n-\deg({\bf G}_1)-1)}{n}$. Hence we have $$R_{storage}+f \leq 1-\frac{g}{n}.$$ The range of the storage rate is $$\frac{g-1}{n} <R_{storage}<1.$$ Another advantage is that ${\bf C} \star {\bf D} \subset {\bf C}={\bf C}(P_1, \ldots, P_n, {\bf G}_1+{\bf G}_2, {\bf X})$. Thus $d({\bf C} \star {\bf D}) -1\geq n-\deg({\bf G}_1)-\deg({\bf G}_2)-1$.  We have the following result.\\

{\bf Theorem 3.1.} {\em Let ${\bf C}$ and ${\bf D}$ are defined as above. Then we have an PIR scheme for ${\bf C}$-coded distributed storage system. The number of tolerated failed server is upper bounded by $d({\bf C})-1=n-\deg({\bf G}_1)-1$. This PIR scheme protects against any $d^{\perp}({\bf D})-1=\deg({\bf G}_2)-2g+1$ colluding servers. The retrieval rate is $R_{retrieval}=\frac{d({\bf C} \star {\bf D})-1}{n}\geq \frac{n-\deg({\bf G}_1)-\deg({\bf G}_2)-1}{n}$. When both ${\bf C}$ and ${\bf C}(P_1, \ldots, P_n, {\bf G}_1+{\bf G}_2, {\bf X})$ are transitive, the rate of the PIR scheme can be improved to $\frac{n-(\deg({\bf G}_1+\deg({\bf G}_2)-g+1)}{n}$.}\\

In the query phase and reconstruction phase the generator matrix of the dual code ${\bf C} \star {\bf D}$ has to be used. In most cases ${\bf C} \star {\bf D}$ is just the functional code ${\bf C}(P_1, \ldots, P_n, {\bf G}_1+{\bf G}_2, {\bf X})$ and the dual is the residual code ${\bf C}_{Omaga}(P_1, \ldots, P_n, {\bf G}, {\bf X})$, see \cite{TV}. The generator matrix of ${\bf C} \star {\bf D}$ can be explicitly given for most curves and AG codes.\\

{\bf Proposition 3.1.} {\em Let ${\bf X}$ be an absolutely irreducible smooth genus $g$ curve defined over ${\bf F}_q$ with the canonical divisor $K$. Let ${\bf G}$ and ${\bf G}'$ be two linear equivalent effective divisors and ${\bf P}=\{P_1, \ldots, P_n\}$ be a set of $n$ rational points satisfying $supp ({\bf G}_i) \bigcap {\bf P}=\emptyset$. Then two functional codes ${\bf C}(P_1, \ldots, P_n, {\bf G}_1, {\bf X})$ and ${\bf C}(P_1, \ldots, P_n, {\bf G}_2, {\bf X})$ are equivalent linear codes. Therefore the dual residual code of ${\bf C}(P_1, \ldots, P_n, {\bf G}, {\bf X})$ is equivalent to a functional code ${\bf C}(P_1, \ldots, P_n, {\bf G}', {\bf X})$, where ${\bf G}'$ is a divisor linear equivalent to $$P_1+\cdots+P_n+K-{\bf G}.$$}\\

{\bf Proof.} Let $f$ be a rational function such that the divisor $(f)$ associated with $f$ is ${\bf G}_1 -{\bf G}_2$. Then the space ${\bf L}({\bf G})=\{g: (g)+{\bf G}_1 \geq 0\}$ of rational functions associated with ${\bf G}_1$ corresponds to ${\bf L}({\bf G}_2)=\{g_1: (g)+{\bf G}_2\geq 0\}$ by $$g \longrightarrow fg=g_1.$$ Then the two functional codes are equivalent.\\

{\bf Corollary 3.1.} {\em Let ${\bf X}$ be an absolutely irreducible smooth genus $g$ curve with $N$ rational point defined over ${\bf F}_q$. Then for a storage rate $R_{storage}$ satisfying $\frac{g-1}{n}<R_{storage}<1-\frac{3g-2}{n}$ AG-coded distributed storage system over $2g-2<n\leq N-1$ servers, and any given $t$ satisfying $$0\leq t \leq 1-R_{storage}-\frac{3g-2}{n},$$  we can construct a star product PIR scheme with the retrieval rate $$R_{retrieval}=1-R_{storage}-t-\frac{3g-2}{n}.$$ This star product PIR scheme protects against any $tn$ colluding servers.}\\

In the case of replicated data storage, that is, $R_{storage}=\frac{1}{n}$, we construct a retrieval rate $$R_{retrieval}=\frac{n-t-3g+1}{n}$$ star product PIR scheme with $t$ colluding servers for the replicated data stored across $n$ servers if $n \leq N-1$. Comparing with the asymptotic capacity $\frac{n-t}{n}$ when the number of the files $m$ goes to the infinity,  see \cite{SunJafar1}, this AG code based star product PIR scheme with colluding servers is good. In this PIR scheme $n$ servers with any given $n$ satisfying $$q^b <n <N-1$$ can be used in the distributed storage system. The elliptic curve case $g=1$ is interesting. In $g=1$ case $N-1=q+\lfloor 2\sqrt{q} \rfloor$ servers can be used in the distributed storage system and the retrieval rate can achieve $$R_{retrieval}=\frac{n-t-2}{n}.$$\\

{\bf Example 1.} There is a genus $6$ curve with $33$ rational points defined over ${\bf F}_8$. Then from Corollary 3.1 for any storage rate $R_{storage}$ coded distributed storage system,  where $$\frac{5}{32}<R_{storage}<\frac{1}{2},$$ and any ratio $t$ of colluding serves satisfying $$t \leq \frac{1}{2}-R_{storage}$$ we can construct a star product PIR scheme with the retrieval rate $$R_{retrieval}=\frac{1}{2}-R_{storage}-t.$$ This PIR scheme can protect against any $32t$ colluding servers.\\

From the main result in \cite{Stich} about transitive AG-code sequence attaining the Tsfasman-Vl\'{a}dut-Zink bound can be constructed from curve sequence in \cite{Garcia}. Thus we have the following result.\\

{\bf Corollary 3.2.} {\em Let ${\bf F}_{q^2}$ is a fixed base field. Let $R_1$ and $R_2$ be two positive real numbers satisfying $R_1+R_2 \geq \frac{1}{2}$. Then we have a sequence of transitive AG-coded distributed storage with the asymptotic storage rate $$R_{asym, storage}=R_1,$$ and $t_i$-privacy PIR schemes with the asymptotic retrieval rate $$R_{asym, retrieval}=1-R_1-R_2-\frac{1}{q-1}$$ and asymptotic colluding ratio $$t_{asym} \geq R_2-\frac{1}{q-1}.$$}\\

\subsection{Colluding servers, Byzantine servers and unresponsive servers}

A Byzantine server in distributed storage system is a server may respond wrong response (error) and an unresponsive server is a server may respond nothing (erasure). In \cite{Taje1,Saarela} star product PIR schemes with colluding, Byzantine and unresponsive servers are constructed from generalized Reed-Solomon codes and binary Reed-Muller codes. Another symbol retriever vector $E$ satisfying items 1)-6) in page 2 of \cite{Saarela} is needed. When both codes are transitive item 6) is satisfied automatically. Then we need find a vector $E \in {\bf F}_{q^b}^n$ such that the linear code $${\bf C_{\star E}^{\star D}}={\bf C} \star {\bf D}+{\bf C} \star {\bf E}$$ can be explicitly given and has its minimum distance at least $2b+a+1$, where $b$ is the number of (changed) Byzantine servers and $a$ is number of (changed) unresponsive servers. When the retrieval code is ${\bf D}={\bf C}(P_1, \ldots, P_n, {\bf G}_2, {\bf X})$ satisfying $$t+2g-1 \leq \deg ({\bf G}_2) \leq n,$$  To decode the responses, $E$ needs to satisfy
$${\bf C_{\star E}^{\star D}} \subset ({\bf C} \star {\bf D})^{\perp},$$ see page 3 of \cite{Saarela}. We use the third functional code ${\bf C}(P_1, \ldots, P_n, {\bf G}_3, {\bf X})$ associated with the the third divisor ${\bf G}_3$ satisfying that $$\deg({\bf G}_1+\deg({\bf G}_2)+\deg({\bf G}_3)+2b+a+1 \leq n,$$ and ${\bf G}_3$ is always linear equivalent to an effective divisor ${\bf G}_3' \leq P_1+\cdots+P_n+K-2{\bf G}_1-2{\bf G}_2$, where $K$ is the canonical divisor of ${\bf X}$. The retrieval vector $E$ is in $${\bf C}(P_1, \ldots, P_n, {\bf G}_1+{\bf G}_2+{\bf G}_3, {\bf X})\setminus {\bf C}(P_1, \ldots, P_n, {\bf G}_1+{\bf G}_2, {\bf X}).$$  Then from Proposition 3.1 ${\bf C_{\star D}^{\star E}}$ is in a suitable equivalent code of $({\bf C} \star {\bf D})^{\perp}$.\\

There are many transitive AG codes. For example, AG codes from elliptic curves with the groups of rational points are of the form ${\bf Z}/N{\bf Z}$, $N$ is the number of rational points, see \cite{Ruck}, are transitive. AG codes from curves in \cite{Garcia} are also transitive. Thus AG code based star product PIR schemes can be constructed from transitive AG codes from some elliptic curves and curves in \cite{Garcia} to protect against colluding, Byzantine and unresponsive servers. Notice that the curves in \cite{Garcia} over ${\bf F}_{q^2}$ attains the Drinfeld-Vl\'{a}dut bound and thus have many rational point. Hence AG-code based star product PIR schemes with colluding, Byzantine and unresponsive servers can be constructed over small fields when large number of servers are used in the distributed storage systems. The retrieval rate can achieve $$\frac{n-\deg({\bf G}_1)-\deg({\bf G}_2)-\deg({\bf G}_3)-1}{n}.$$

For example for an elliptic curve defined over ${\bf F}_{2^s}$ with $N$ rational points, such that the group of rational points of this curve is of the form ${\bf Z}/N{\bf Z}$. If $N$ is a prime number, then codes associated with $mO$ and evaluated at all other $N-1$ nonzero elements are transitive, where $O$ is the zero element of the group of all rational points and $P_1, \ldots, P_{N-1}$ are other $N-1$ nonzero elements of the group. Notice that $P_1+\cdots+P_{N-1}$ is linear equivalent to the divisor $(N-1)O$. The canonical divisor of an elliptic curve is always zero. The three divisors ${\bf G}_i$ are of the form $m_iO$, where $m_1, m_2, m_3$ are three positive integers satisfying $$2g-2<m_i <N-1,$$ $$t+1 \leq m_2,$$  $$2b+a+1 \leq n-m_1-m_2-m_3,$$ $$2(m_1+m_2)+m_3 \leq N-1.$$ Set $${\bf C}={\bf C}(P_1, \ldots, P_n, {\bf G}_1, {\bf X}),$$ and $${\bf D}={\bf C}(P_1, \ldots, P_n, {\bf G}_2, {\bf X}).$$ Then symbol retrieval vector $E$ is in the code ${\bf C}(P_1, \ldots, P_n, {\bf G}_1+{\bf G}_2+{\bf G}_3, {\bf X})$. Then ${\bf  C_{\star E}^{\star D}} \subset $ ${\bf C}(P_1, \ldots, P_n, {\bf G}_1+{\bf G}_2+{\bf G}_3, {\bf X})$ $ \subset {\bf C}(P_1, \ldots, P_n, {\bf G}_1+{\bf G}_2, {\bf X})^{\perp}$. Then we construct an elliptic curve code based star product PIR scheme protecting against $t$ colluding. $b$ Byzantine, $a$ unresponsive servers. The retrieval rate is $$R_{retrieval}=\frac{N-1-m_1-m_2-m_3-1}{N-1}.$$

In many cases $N$ is bigger than $2^s$ and this kind of elliptic curve code star product PIR schemes can be constructed over the field ${\bf F}_{2^s}$ when $N-1>2^s$ servers are used in the distributed storage system.\\

\section{Cyclic code based star product PIR schemes for replicated data}

PIR schemes using cyclic codes were mentioned in \cite{Freij1,KLRA,Saarela}. In \cite{BMR21} PIR schemes with colluding servers using cyclic codes were studied. In particular in the case of $2^m-1$ server distributed storage system, parameters of cyclic code based PIR schemes with colluding servers from length $2^m-1$ BCH codes are compared with punctured Reed-Muller code based PIR schemes with colluding servers.\\

Notice that the number of servers can be arbitrary positive integer $n$ as showed in this section, and the number of colluding servers can be more flexible in binary cyclic code based star product PIR schemes with colluding servers than star product PIR schemes for Reed-Muller coded distributed storage studied in \cite{Freij1,Saarela}. In this section we consider the replicated storage of data, that is, the storage code ${\bf C}$ is the $[q^m-1, 1, q^m-1]_q$ code. Let ${\bf D}$ be a primitive length $q^m-1$ narrow-sense BCH code over ${\bf F}_q$ studied in Theorem 18 ($q=2$) or Theorem 23 ($q \geq 3$) in \cite{GDL}. It is clear that all cyclic codes are transitive. Then the rate is always $\frac{\dim({\bf D}^{\perp})}{n}$ and the number of colluding servers is at most $d^{\perp}({\bf D})-1$.\\

When $q=2$ set $r=m-t$ where $2 \leq t \leq m-1$, $$k_t=2^m-1-\Sigma_{i=1}^{\lfloor \frac{m}{r+1} \rfloor} (-1)^{i-1} \frac{m}{i} \displaystyle{m-ir-1 \choose i-1} \cdot 2^{m-i(r+1)}.$$ Then we have a binary dually BCH code ${\bf D}^{\perp}$ (the dual is also BCH) with dimension $k$ in the range $$2^m-1-k_t\leq k <2^m-1-k_{t+1}$$ and distance $d \geq 2^{m-t}$, where $2 \leq t \leq m-3$ from Theorem 18 of \cite{GDL}. Set the retrieval code ${\bf D}$ as above we have a binary BCH code based star product PIR scheme for replicated data storage which protects against $2^{m-t}-1$ colluding servers and has the retrieval rate $$\frac{2^m-1-k_t}{2^m-1} \leq R_{retrieval}=\frac{k}{2^m-1} <\frac{2^m-1-k_{t+1}}{2^m-1}.$$

If the number of colluding servers is at most $2^{m-1}-1$, then the retrieval rate is $\frac{m}{2^m-1}$. If the the number of the colluding servers is at most $3$, the retrieval rate is $\frac{2^m-m-2}{2^m-1}$. Comparing with the PIR scheme for replicated data from binary Reed-Muller code in \cite{Freij1}, the retrieval rate is $\frac{1+m+\frac{m(m-1)}{2}}{2^m}$ for $t=2^{m-1}-1$ and the retrieval rate is $\frac{2^m-1}{2^m}$ when $t=3$. Considering the number of servers is $2^m-1$, the binary BCH code based star product PIR schemes are comparable with these PIR schemes from binary Reed-Muller codes.\\

When $q\geq 3$ set $r=m-s$, set $$k_s=q^m-1-\Sigma_{i=1}^{\lfloor \frac{m}{r+1} \rfloor} (-1)^{i-1} \frac{m(q-1)^i}{i} \displaystyle{m-ir-1 \choose i-1} \cdot q^{m-i(r+1)}.$$ Then we have a $q$-ary dually BCH code ${\bf D}^{\perp}$ with dimension $k$ in the range $$q^m-1-k_s\leq k <q^m-1-k_{s+1}$$ and distance $d \geq q^{m-s-1}-q+2$, where $2 \leq s \leq m-3$, see Theorem 23 of \cite{GDL}. Set the retrieval code ${\bf D}$ as above we have a $q$-ary BCH code based star product PIR scheme for replicated data storage which protects against $q^{m-s-1}-q+1$ colluding servers and has the retrieval rate at least $\frac{q^m-1-k_s}{q^m-1}$. If the number of colluding servers is at most $q^m-q^{m-1}-1$, then the retrieval rate is $\frac{m}{q^m-1}$. If the the number of the colluding servers is at most $2$, the retrieval rate is $\frac{q^m-m-2}{q^m-1}$, see Theorem 23 of \cite{GDL}. \\

These $q$-ary BCH code based star product PIR schemes are suitable when $q^m-1$ servers are used in the replicated data storage of $q$-ary files. Comparing with Reed-Muller code based star product PIR schemes with colluding servers in \cite{Freij1,Saarela}, the numbers of colluding servers are flexible.\\

The another interesting case is the number of servers in the replicated storage of data is $q^m+1$. In this case length $q^m+1$ BCH codes are suitable for star product construction. From Corollary 19 about binary BCH codes in \cite{LDL}, we have a star product PIR scheme protecting against at most $10-1=9$ colluding servers and with the retrieval rate $R_{retrieval}=\frac{2^m-4m}{2^m+1}$, or an PIR scheme protecting against at most $14-1=13$ colluding servers and with the retrieval rate $R_{retrieval}=\frac{2^m-6m}{2^m+1}$, or a an PIR scheme protecting against at most $18-1=17$ colluding servers and with the retrieval rate $R_{retrieval}=\frac{2^m-8m}{2^m+1}$.\\

In general we have the following result from Theorem 18 in \cite{LDL}.\\

{\bf Theorem 4.1.} {\em Let $q$ be a prime power $\delta$ be a positive integer satisfying $3 \leq \delta \leq q^{\lfloor \frac{m-1}{2} \rfloor }+3$. We consider the PIR scheme over ${\bf F}_q$ with colluding servers for replicated storage of data. Then we have an BCH code based star product PIR scheme protecting against $2\delta-3$ colluding servers. The retrieval rata of this PIR scheme is $$R_{retrieval}=\frac{q^m-2m(\delta-2-\lfloor \frac{\delta-2}{q} \rfloor)}{q^m+1}.$$}\\

Now we consider the replicated data storage across $3 \cdot 2^s$ servers. This case can be comparing with Reed-Muller code based star product PIR schemes across $2^s$ servers. Then length $3 \cdot 2^s$ repeated-root cyclic codes studied in \cite{LiuLi} can be used. Let $s$ be a fixed positive integer and $s_0,s_1$ be two nonnegative integers satisfying $0\leq s_1\leq s_0 \leq s-1$. If $i$ and $j$ are two positive integers satisfying $$2^s-2^{s-s_0}+1 \leq i \leq 2^s-2^{s-s_0}+2^{s-s_0-1},$$ $$2^s-2^{s-s_1}+1 \leq i \leq 2^s-2^{s-s_0}+2^{s-s_1-1},$$ then the binary repeated-root cyclic code generated by $(x+1)^i(x^2+x+1)^j$ has the dimension $3 \cdot 2^s-i-2j$ and the minimum Hamming distance $\min\{2^{s_0+1}, 2^{s_1+2}\}$, see \cite{LiuLi} Table 1 and Table 2. Thus we have the following result.\\

{\bf Theorem 4.2.} {\em Let $s, s_0, s_1, i, j$ be nonnegative integers as above. We consider the PIR scheme over ${\bf F}_2$ with colluding servers for replicated storage of data. Then we have a repeated-root cyclic code based star product PIR scheme protecting against $\min\{2^{s_0+1}, 2^{s_1+2}\}-1$ colluding servers. The retrieval rata of this PIR scheme is $$R_{retrieval}=\frac{3 \cdot 2^s-i-2j}{3 \cdot 2^s}.$$}\\

\section{Comparison with RS-coded databases and RM-coded databases}
The PIR scheme with colluding servers from Reed-Muller codes were studied in \cite{Freij1,Saarela}. These $t$-privacy PIR for coded database can be defined over the smallest field ${\bf F}_2$ and have good performance. $m$, $r$ and $r'$ are three positive integers satisfying $r+r' \leq m$. The parameters of their PIR schemes are as follows,\\
$$R_{retrieval}=\frac{2^{m-r-r'}-1}{2^m},$$ for PIR scheme in Theorem 2 in \cite{Freij1}, or $$R_{retrieval}=\frac{\Sigma_{i=0}^{m-r-r'} \displaystyle{m \choose i}}{2^m},$$ for PIR scheme in Theorem 4 of \cite{Freij1}, $$t=\frac{2^{r'+1}-1}{2^m},$$ and $$R_{storage}=\frac{\Sigma_{i=0}^r \displaystyle{m \choose i}}{2^m}.$$ It is easy to verify that $1-R_{retrieval}-R_{storage}-t$ is much bigger than $0$.\\

{\bf Example 2.} Let the storage code is the Reed-Muller code $RM(7, 0)$. This is the replicated data storage with $r=0$. Let $r'=3$ and $RM(7, 4)$ be the retrieval code. Then the star product PIR scheme with colluding servers protects against $16-1=15$ colluding servers. The retrieval rate is $\frac{\dim(RM(7, 3))}{2^7}=\frac{1}{2}$. On the other hand there is an extended BCH $[128, 64, 22]_2$ code. Let the retrieval code be the dual of this BCH code. We have a star product PIR scheme with colluding servers protects against $22-1=21$ servers. The retrieval rate is $\frac{1}{2}$. \\

The above example shows a simple fact that when the storage code is Reed-Muller code, the best choice of the retrieval code is not the Reed-Muller code. We will study this problem further in \cite{Chen1}.\\

The PIR scheme with colluding servers from the generalized Reed-Solomon codes were studied in \cite{Freij}. These $t$-privacy PIR for MDS-coded database can only be defined over the field ${\bf F}_q^{b \times k}$ satisfying $n \leq q^b$, that is, each file has at least $log_2n$. Let ${\bf C}$ be a generalized Reed-Solomon $[n, k_1, n-k_1+1]_q$ code and ${\bf D}$ is another generalized Reed-Solomon $[n, k_2, n-k_2+1]_q$ code. Therefore the condition $n\leq q^b$ is needed. Each file ${\bf x}_i \in {\bf F}_q^{b \times k}$ has at least $log_2n$ bits. The parameters of their PIR schemes are as follows,\\
$$R_{retreieval}=\frac{n-k_1-k_2+1}{n},$$ and $$t=\frac{k_2}{n},$$   and $$R_{storage}=\frac{k_1}{n}.$$ Then the PIR scheme for MDS-coded distributed storage attains the Singleton type bound. However only $n \leq q^b$ servers can be used in the distributed storage system.\\

If binary AG codes or AG-cdoes over small fields ${\bf F}_q$, $q \leq 4$, are used, we need the table in \cite{Geer}. Parameters of AG code based star product PIR schemes over ${\bf F}_2$ are worse than PIR scheme from binary Reed-Muller codes. However the number of servers in AG-code based PIR are flexible, not restricted to $2^m$. When the base field is ${\bf F}_q$, $q\geq 8$, star product PIR scheme against colluding servers from AG codes defined over ${\bf F}_8$ have better parameters than parameters of PIR schemes from Reed-Muller codes. More importantly the number of servers and the number of colluding servers are flexible and thus more suitable in practice.\\

\section{Conclusion}

PIR schemes with colluding servers defined over small fields are more suitable in practice. In this paper the Singleton type upper bound for star product PIR schemes with colluding servers for coded distributed storage systems of data from general linear codes was proposed and proved. We studied PIR schemes with colluding servers for AG-coded distributed storage, which can be defined over small fields. Moreover the number of servers and the number of colluding servers in PIR schemes for AG-coded distributed storage are flexible, not restricted as the PIR schemes for MDS-coded databases or PIR schemes for Reed-Muller coded databases. It was proved that parameters in PIR schemes with colluding servers for AG-coded distributed storage are closing to the Singleton type upper bound when the base field is large. We argued that transitive AG-code based star product PIR schemes over small fields can be constructed which protect against colluding servers, Byzantine servers and unresponsive servers, when large number of servers are used in the distributed storage systems. Hence it seems that AG-code based star product PIR schemes over small fields are more suitable in practice. $q$-ary cyclic code based star product PIR schemes with colluding servers for replicated data storage are also studied and the retrieval rates are calculated for various numbers of colluding servers. It showed that when the storage code is the Reed-Muller code, the best choice of the retrieval code is not always the Reed-Muller code.\\

\end{document}